\documentclass[aps,twocolumn,pra]{revtex4-1}

\usepackage{newlfont}
\usepackage{color}         % color to the text \textcolor{red,green,blue}{text to be colored}
\usepackage{soul}            % strickthrough the text \st{text to be stricked} and \underline{text to be underlined}
\usepackage{amssymb}
\usepackage{amsfonts}
\usepackage{amsmath}
\usepackage{wasysym}
\usepackage{graphicx}
\usepackage{epsfig}
\usepackage{amsthm}
\usepackage{bm}
\usepackage{bbold}
\usepackage{mathrsfs}

\usepackage[colorlinks=true, citecolor=blue, urlcolor=blue ]{hyperref}

\newcommand{\ket}[1]{| {#1} \rangle}

\newcommand{\tr}{\textrm{Tr}}

\begin{document}

\definecolor{red}{rgb}{1,0,0}

%\title{All Multipartite Quantum Correlated States are Not Dense Codeable}
\title{
%All Multipartite Quantum Correlated States are Not Good for Multipartite Classical Information Transmission
Multipartite Dense Coding vs. Quantum Correlation: \\
Noise Inverts Relative Capability of Information Transfer
}

\author{Tamoghna Das, R. Prabhu, Aditi Sen(De), and Ujjwal Sen}

\affiliation{Harish-Chandra Research Institute, Chhatnag Road, Jhunsi, Allahabad 211 019, India}

\begin{abstract}
A highly entangled bipartite quantum state is more advantageous for the quantum dense coding protocol than states with low entanglement. Such a correspondence, however, does not exist even for pure quantum states in the multipartite domain. We establish a connection between the multiparty capacity of classical information transmission in quantum dense coding and several multipartite quantum correlation measures of the shared state, used in the dense coding protocol. In particular, we show that for the noiseless channel, if multipartite quantum correlations of an arbitrary multipartite state of arbitrary number of qubits is the same as that of the corresponding generalized Greenberger-Horne-Zeilinger state, then the multipartite dense coding capability of former is the same or better than that of the generalized Greenberger-Horne-Zeilinger state. Interestingly, in a noisy channel scenario, where we consider both uncorrelated and correlated noise models, the relative abilities of the quantum channels to transfer classical information can get inverted by administering a sufficient amount of noise. When the shared state is an arbitrary multipartite mixed state, we also establish a link between the classical capacity for the noiseless case and multipartite quantum correlation measures.
\end{abstract}

\maketitle

\section{Introduction}

In recent times, a lot of interest has been created to characterize and quantify quantum correlations in multipartite quantum systems \cite{HoroRMP,DiscRMP,USAdvPhy,ManybodyRMP}. This is due to the fact that the preparation of multiparticle states with quantum coherence enables us to realize several quantum information protocols like quantum dense coding \cite{DCref}, quantum teleportation \cite{TeleRef}, secure quantum cryptography \cite{QCryp}, and one way quantum computation \cite{onewayQC}, in a way that is better than their classical counterparts. This increasing interest is further boosted by the latest advances in experiments to realize multipartite states in different physical systems including photons, ion traps, optical lattices, and nuclear magnetic resonances \cite{Pan10photon, Blatt14ion, Blatt6ion, Mahesh12NMR, Bloch, Bosons}. 

It has been shown, both theoretically as well as experimentally, that bipartite entanglement is an essential ingredient for a vast majority of known quantum communication schemes, involving two parties.  Specifically, it has been established that in the case of  pure bipartite states,  the capacity of classical information transmission via quantum states increases with the increase of any quantum correlation measure. Such a simple scenario is facilitated by the fact that quantum correlation in a bipartite pure quantum state, in the asymptotic domain, can be {\em uniquely} quantified by  the local von Neumann entropy \cite{HoroRMP}. The situation is far richer in the multiparty case (see Refs. \cite{qwe1,qwe2,qwe3,qwe4,qwe5,qwe6,qwe7,qwe8,qwe9,qwe10}).
%\textcolor{red}{Specifically, in the multipartite case, there exists various classes of states with various degree of entanglement \cite{richmulti}. For example, in case of three-qubit states there are three classes -- fully separable states, biseparable states, and genuine multipartite entangled states \cite{3qubitclasses}.} 
%
%In particular, for three qubit states there are 3 types of states, full , bi and tri seperable. in perticular there will be n diffienrt classes like fully bi tri etc. 
%{\bf examples and refs - - -. }  %which helps to obtain such a relation. However, it is not the case in a multipartite scenario.  Depending on the entanglement content, 
Regarding the entanglement content, in a multiparty quantum state, which ranges from being bipartite to genuine multipartite, even multipartite {\em pure} states can be classified in several ways \cite{3qubitclasses} and therefore, there is no unique measure that can determine quantum correlations, present in the system. 

Unlike point to point communication, where among two parties, one acts as a sender and the other as a receiver, communication protocols involving multiple parties can have various complexities. 
%One of the possible setting can be a single sender and many receivers in which sender sends individually information to all the recievers. 
%Hence, such schemes, involving multipartite systems play a vital role in the developments of information theory. It has been shown that quantum correlation present in the system is required to realize many such communication protocols in  more efficient ways than their classical counterparts. Therefore, is important to find the link between the capability of the classical information transmission via quantum states and the 
% multipartite classical information transfer and 
%quantum correlation measures.
One possible scenario involves  several senders and a single receiver. 
%
%Such interdependence is possible partially because quantum correlations shared between arbitrary two parties can be characterized and quantified \cite{HoroRMP}. 
%one-to-one-correspondence is possible partially because quantum correlations  shared between two parties are well understood \cite{HoroRMP}. In contrast,
%%However in a multipartite scenario, the question of usefulness of entanglement in quantum communication as well as computational tasks  is unanswered \cite{HoroRMP,DiscRMP}. For addressing such question, there are several obstacles like  quantification of multipartite entanglement even for pure states  is not unique due to the existing  hierarchies of mutlpartite states according to their content of quantum correlations, and
%the quantification of multipartite quantum correlations is not unique even for tripartite pure states and hence .
%At the same time, %finding the capacity of information transmission involving many parties are also restricted due to their complex nature \cite{physnews}. 
%In bipartite domain, information transfer involves only two parties among which one acts as a sender and another  as a receiver. In the multipartite communication setting, one of the possible scenario involves
%Such a situation happens,  for example,    
Suitable examples for such multipartite communication protocols include, when several  news reporters from different locations send various news articles to the newspaper editorial office or when several weather observers from different places communicate their respective weather reports to the regional meteorological office. 

In this paper, we connect multipartite communication protocols with multiparty quantum correlation measures. In particular, we establish a relation between the capacity of multipartite dense coding and multipartite quantum correlation measures of arbitrary multiqubit states. This correspondence is illustrated both in the case of noiseless and  noisy channels for arbitrary shared states. 
%Unlike bipartite scenario, the capacity of classical information transfer and multipartite quantum correlation are not bijective, 
%multipartite states does not always have one-to-one correspondence with the capacity of classical information transfer.  
Specifically, we show that in the case of the noiseless quantum channel,  if the capacity of classical information transmission using the generalized Greenberger-Horne-Zeilinger ($g$GHZ) state \cite{GHZ} is the same as that of any  multiqubit pure state, then the multipartite quantum correlation
%of those states and the gGHZ state are related in two ways:
%the relation between multipartite quantum correlations of arbitrary states and the correlations of the gGHZ state are of two kinds -- either  the multiparty quantum correlations
of the $g$GHZ state is either the same or higher than that of the latter states. The result is generic in the sense that it does not rely on the choice of the quantum correlation measure, and is independent of the number of parties. Three computable measures, the generalized geometric measure \cite{qwe10,GGM,GM}, the  tangle \cite{CKW} and discord monogamy score \cite{ourdiscmono,discscore}, are considered for obtaining the results.
%the gGHZ state and those states have equal multipartite quantum correlation or  arbitrary state has less  multipartite quantum correlation than that of the gGHZ state.  
It is to be noted that while the first multiparty measure is based on the geometry of the space of multiparty quantum states, the latter two are based on the concept of monogamy of quantum correlations.
%When the shared state is an arbitrary multipartite mixed state,  we obtain a sufficient condition  for dense codeablity. If  arbitrary three-qubit mixed states of rank-2 are shared between two senders and a single receiver, numerical simulations indicate that the relation obtained for entanglement and capacities of arbitrary pure states and that of the gGHZ state does not hold. We then show that in the presence of moderate noise, one can  change the ordering between the quantum correlations of a set of arbitrary state and that of the gGHZ state with respect to the dense coding capability. We find that the effects of both correlated and uncorrelated covariant noise on the capacities of the gGHZ state is much less compared to any arbitrary three qubit pure states. Therefore, with respect to the transmission of classical information, the gGHZ state  is  more robust than  the arbitrary three-qubit pure states. In all the cases, the quantum correlation of arbitrary states is measured before encoding.  
%This leads to a classification of arbitrary three party states, according to their advantages in dense coding protocol, in presence of noise.   
%It leads to a classification of 

The noisy case can be considered at several levels of complexity. Here we consider the following two situations. First, we consider the case when the noise is present before the encoding of classical information, while the quantum channel that transmits the encoded state is noiseless. Secondly, we consider the situation when the encoding is performed on a pure shared state, while the post-encoded state is sent via a noisy quantum channel. In the first case, we begin by obtaining a sufficient condition for dense codeability for arbitrary, possibly mixed, multipartite quantum states. We perform numerical simulations by generating Haar uniform rank-2 three-qubit states, and find that a great majority of them are better carriers of classical information than the $g$GHZ state with the same multiparty quantum correlation content.  Numerical analysis of higher-rank states are also considered.

\begin{figure}[t]
\begin{center}
 \includegraphics[width=1\columnwidth,keepaspectratio,angle=0]{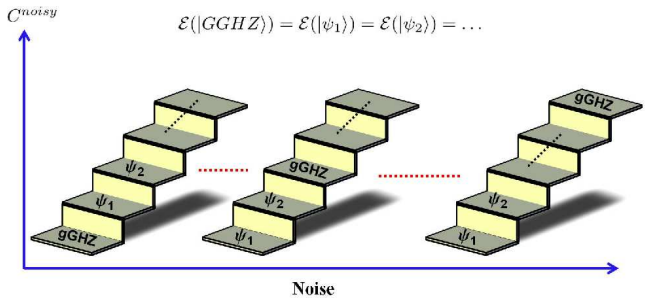}
 \end{center}
\caption{Schematic diagram of the change of status of the $g$GHZ state in comparison to other multipartite states with respect to multipartite DC capacity in the presence of noise. The comparison has been made 
%between DC capacity of the gGHZ state and that of the arbitrary states, 
with the states which posses same amount of multipartite quantum correlations as the $g$GHZ state. The results obtained in this paper shows that the $g$GHZ state is more robust against noise as compared to arbitrary states for the dense coding protocol. This is independent of the fact whether the noise in the system is from the source or in the channel after the encoding.}
\label{fig:Staircase}
 \end{figure}

Going over to the second case, when the quantum channel is noisy, we show that the presence of noise can invert the relative capability of information transfer for two states with the same multiparty quantum correlation content.
%In case of arbitrary three-qubit states, containing same amount of  quantum correlation, a classification of multiparty states, based on the quantum advantages in the dense coding protocols  with noisy environements can be obtained (See Fig. 1). 
In particular, we find that the effect of both correlated and uncorrelated covariant noise with respect to the capability of classical information transfer is less pronounced on the $g$GHZ state as compared to a generic pure multiparty state. This relative suppression of the effect of noise for the $g$GHZ state is what results in the inversion of the relative capabilities of information transfer between a $g$GHZ state and a generic pure multiparty quantum state. A schematic diagram elucidating the situation is presented in Fig. \ref{fig:Staircase}.

%The classification can be schematically seen in Fig. 1.    

%case of moderate noise in the channel, specifically for correlated Pauli channel, we find that the  set of states for which  capacities  as well as genuine multipartite entanglement has one to one correspondence with the $g$GHZ state for the noiseless case,  require more genuinely multipartite entanglement to make themselves equal to the capacities of the $g$GHZ state. Therefore, the presence of noise makes the  $g$GHZ state superior %with respect to 
%in the capability of dense coding protocol  among the set of multiqubit pure states, having same amount of  multipartite entanglement.} 

The paper is organized as follows. In Sec. \ref{sec:QCM&M}, we describe the quantum correlation measures that are used later in the paper. In the subsequent section (Sec. \ref{sec:QDC}), we discuss the capacities of the quantum dense coding protocol, with and without noise. In Sec. \ref{sec:relationsnoiselss}, we establish the connection between the capacity and the quantum correlation measures for the noiseless channel, while  we deal with the noisy channels in Sec. \ref{sec:relationsnoisy}. In particular, we  deal with the fully correlated Pauli channel in Sec. \ref{subsec:fullycorrelatedPauli} and  with the uncorrelated Pauli channel in Sec. \ref{subsec:uncorrelatedPauli}. We present a conclusion in Sec. \ref{sec:conclusion}.

\section{Quantum correlation measures}
\label{sec:QCM&M}

Quantum correlations present in bipartite quantum systems can broadly be classified into two paradigms -- (a) entanglement-separability  and (b) information-theoretic. Measures of the former paradigm  always vanish for separable states.
% are characterized  by being vanishing for separable states. 
Examples include entanglement of formation \cite{distillable}, concurrence \cite{concEof},  distillable entanglement \cite{distillable}, and logarithmic negativity \cite{LN}. On the other hand, information-theoretic quantum correlation measures are independent of entanglement, and examples include quantum discord \cite{discord} and quantum work deficit \cite{workdeficit}.
%etc., belong to the information-theoretic paradigm. The later ones are independent from the concepts of the entanglement.
%Information-theoretic quantum correlation measures have been coined to quantify the entanglement free quantum correlations present in systems. 
In the present section, we define concurrence and quantum discord, and then discuss monogamy relations based on these two measures. Finally we define a multipartite entanglement measure, the generalized geometric measure, based on the concept of the Fubini-study metric  \cite{qwe10,GGM} (cf. \cite{GM}). 
% manifest clearly. One should note that monogamy relations for entanglement measures are always monogamous, where as, the information theoretic measures can also be non-monogamous \cite{ourdiscmono,Giorgi,Fanchini,Salini}.

\subsection{Concurrence}
\label{subsec:C&T}

Concurrence is a quantum correlation measure for two-qubit systems, which is a monotonic function of entanglement of formation \cite{concEof}. 
%Using concurrence one can quantify quantum correlation for any two-qubit state. 
For an arbitrary two-qubit state, $\rho_{AB}$, it  is given by 
\begin{equation}
{\cal C}(\rho_{AB})=\textrm{max} \{0,\, \lambda_1-\lambda_2-\lambda_3-\lambda_4 \},
\end{equation}
where the $\lambda_i$'s are the square roots of the eigenvalues of  $\rho_{AB}\tilde\rho_{AB}$ in decreasing order and 
$\tilde\rho_{AB}=(\sigma_y \otimes \sigma_y) \rho_{AB}^* (\sigma_y \otimes \sigma_y)$, with  the complex conjugate being taken in the computational basis. 
%of the state. For any pure two-qubit state, $|\psi_{AB}\rangle$, the concurrence is given by $2\sqrt{\textrm{det}\rho_A}$, where $\rho_A$ is the marginal density matrix of the bipartite state $|\psi_{AB}\rangle$.
Concurrence vanishes for all separable states while it is maximal for any maximally entangled state.

%In the monogamy score relation, see Eq. (\ref{Eq:monoQ}), 

\subsection{Quantum Discord}
\label{subsec:D&DMS}

Quantum discord is an information-theoretic quantum correlation measure, which is obtained by taking the difference between two inequivalent forms of quantum mutual information.  Mutual information quantifies the correlation between two systems. Classically, it can be defined in two equivalent ways. 
%(i) If the lack of information is measured by Shannon entropy, $H(X)=-\sum_x p_x \log p_x$, where $p_x$ is the probability of $x$ occurring as a value for the classical variable $X$,  then
%In classical information theory,  mutual information quantifies the correlation between 
For two variables, $X$ and $Y$, it  is defined as
\begin{equation}
\label{Eq:MIi}
I(X,Y)=H(X)+H(Y)-H(X,Y),
\end{equation}
where $H(X)=-\sum_x p_x \log_2 p_x$ is the Shannon entropy, with $p_x$  being the probability of $x$ occurring as a value for the classical variable $X$, and similarly for $H(Y)$. 
 $H(X,Y)$ denotes the Shannon entropy of the joint probability distribution of $X$ and $Y$. Using Bayes' rule, one can rewrite the  mutual information in terms of conditional entropy, $H(X|Y)$, to obtain
 \begin{equation}
  I(X,Y)= H(X)- H(X|Y).
 \end{equation}

%ll logarithms are with base two.  The above equation gives the difference between the information content in composite system and the addition of information in subsystems.
%(ii) Using Bayes' rule one can write the conditional probability as $p_{x|y}=p_{xy}/y$, hence the equivalent mutual information takes the form
%\begin{equation}
%\label{Eq:MIii}
%J(X,Y)=H(X)-H(X|Y)
%\end{equation}
%where $H(X|Y)$ is the conditional entropy. The above equation gives the gain in information about one subsystem due to the measurement on the other.

%Under quantum mechanical considerations, 

In the quantum domain, these two classically equivalent definitions of  mutual information become unequal and their difference has been proposed to be  a measure of quantum correlation, called the quantum discord \cite{winterPRA,discord}.
% Quantum version of the above definitions: (i) Replacing the Shannon entropies by von Neumman entropies in Eq. (\ref{Eq:MIi}), 
For any composite system, $\rho_{AB}$, quantizing the first definition of the mutual information, one obtains
\begin{equation}
\label{Eq:MIqi}
{\cal I}(\rho_{AB})=S(\rho_A)+S(\rho_B)-S(\rho_{AB}),
\end{equation}
where $S(\varrho)=-\tr (\varrho \log \varrho)$ is the von Neumann entropy of $\varrho$. This quantity has been argued to be total correlation in the bipartite state \cite{discord}.
Quantizing the second definition is not straightforward, since the quantity obtained by replacing the Shannon entropies by the von Neumann ones can be negative for some quantum states \cite{CerfAdami}. To overcome this drawback, one can make a measurement on one of the subsystems, say subsystem $B$, of  $\rho_{AB}$, and the measured conditional entropy of $\rho_{AB}$ can be obtained as
\begin{equation}
\label{Eq:CondEntropy}
S(\rho_{A|B})=\min_{\{B_i\}}\sum_i p_i S(\rho_{A|i}),
\end{equation}
where the rank-1 projection-valued measurement $\{B_i\}$ is performed on the $B$-part of the system. Here $\rho_{A|i}=(1/p_i)(\tr_B[(\mathbb{1}_A \otimes B_i) \rho_{AB} (\mathbb{1}_A \otimes B_i)])$, with $p_i=\tr_{AB}[(\mathbb{1}_A \otimes B_i) \rho_{AB} (\mathbb{1}_A \otimes B_i)]$, and $\mathbb{1}_A$ being the identity operator on the Hilbert space of subsystem $A$. Using this quantity, one then quantizes the second definition of mutual information as 
\begin{equation}
\label{Eq:MIqii}
{\cal J}(\rho_{AB})=S(\rho_A)-S(\rho_{A|B}),
\end{equation}
which has been argued to be a measure of classical correlation of the bipartite state \cite{discord}.
Finally, the quantum discord is defined as
%the difference between two inequivalent mutual information's given in Eqs. (\ref{Eq:MIqi}) and (\ref{Eq:MIqii}), i.e.,
\begin{equation}
\label{Eq:QDiscord}
{\cal D}(\rho_{AB}) = {\cal I}(\rho_{AB}) - {\cal J}(\rho_{AB}).
\end{equation}
%Discord vanishes only if the system $\rho_{AB}$ is classical quantum.

\subsection{Monogamy score: Tangle and Discord Monogamy Score}

Monogamy of quantum correlations quantifies the sharability of the same in  multipartite systems \cite{CKW}.  
For an arbitrary $(N+1)$-party quantum  state, $\rho_{A_1A_2 \ldots A_NB}$,
%, shared between three parties viz., Alice($A$), Bob($B$), and Charu($C$). 
let ${\cal Q}_{A_iB}$ ($i=2,\ldots, N$)  be the amount of a certain quantum correlation shared between the pair $A_iB$ ($i=2,\ldots, N$), and 
 ${\cal Q}_{A_1A_2 \ldots A_N:B}$  represent the same between  $B$ and rest of the parties. Here  $B$ acts as a ``nodal'' observer. The state $\rho_{A_1A_2 \ldots A_NB}$ is said to be monogamous for the quantum correlation measure, ${\cal Q}$, if \cite{CKW}
%if ${\cal Q}_{AB} + {\cal Q}_{AC}$ is bounded by ${\cal Q}_{A:BC}$, i.e., for a state to be monogamous the following inequality holds,
\begin{equation}
\sum_{i=1}^{N} {\cal Q}_{A_iB}  \leq {\cal Q}_{A_1A_2 \ldots A_N:B}.
\end{equation}
Using the terms of the above relation, we define the monogamy score for the quantum correlation measure, ${\cal Q}$, as
\begin{equation}
\label{Eq:monoQ}
\delta_{{\cal Q}}= {\cal Q}_{A_1A_2 \ldots A_N:B} -\sum_{i=1}^{N} {\cal Q}_{A_iB},
\end{equation}
for $B$ as the nodal observer. Therefore, ${\cal Q}$  is monogamous for a given state when 
$\delta_{\cal Q}$ is positive for that state. Otherwise, the measure is said to be non-monogamous for that state. The advantage of such a multiparty quantum correlation measure  is that it can be expressed in terms of bipartite quantum correlation measures.
%Since for two spin-$\frac{1}{2}$ particles, quantum correlation measures are well-understood,  the monogamy based measures are computable for arbitrary multiqubit states.

In Eq. (\ref{Eq:monoQ}), if ${\cal Q}$ is chosen to be  the square of the concurrence, then we obtain the tangle \cite{CKW}, which is known to be monogamous for all multiqubit states \cite{CKW,FVTO}. Choosing ${\cal Q}$ to be quantum discord, we obtain the discord monogamy score \cite{discscore}, which can be negative even for some three qubit pure states \cite{ourdiscmono}.
%for $(N+1)$-party state $\rho_{A_1 A_2 \ldots A_NB}$, which is given by 
%\begin{equation}
%\label{Eq:tangle}
%\delta_{{\cal C}^2}= {\cal C}^2_{A_1A_2 \ldots A_N:B} -\sum_{i=1}^{N} {\cal C}^2_{A_iB}  .
%\end{equation}
% and is given by
%\begin{equation}
%\label{Eq:tangle}
%\delta_{\tau}(\rho_{ABC})={\cal C}^2(\rho_{A:BC})-{\cal C}^2(\rho_{AB})-{\cal C}^2(\rho_{AC}).
%\end{equation} 
%Note that the tangle is invariant under permutation of particles. In other words, to evaluate tangle it is irrelevant which one of the three parties is chosen as a nodal observer. For any genuinely entangled state, tangle will vanish.
 %For arbitrary number of parties, it is shown that the tangle is positive definite and hence it is monogamous for all states.

%On the other hand, the discord monogamy score is obtained by considering quantum discord as the quantum correlation measure and hence the discord monogamy score is
%\begin{equation}
%\label{Eq:discomono}
%\delta_{{\cal D}}= {\cal D}_{A_1A_2 \ldots A_N:B} -\sum_{i=1}^{N} {\cal D}_{A_iB}  .
%\end{equation}
%Discord monogamy score can be negative for some three-qubit pure states \cite{ourdiscmono}. 

\subsection{Generalized Geometric Measure}
\label{subsec:GGM}

Let us now define a genuine multipartite entanglement measure. An $N$-party pure state is said to be genuinely multiparty entangled if it is non-separable with respect to every bipartition. The generalized geometric measure (GGM) is obtained by considering the minimal distance from the set of all multiparty states that are not genuinely multiparty entangled \cite{qwe10,GGM} (cf. \cite{GM}).
More specifically, the GGM of  an \(N\)-party pure quantum state \(|\phi_N\rangle\) is defined as
\begin{equation}
{\cal E} ( |\phi_N\rangle ) = 1 - \Lambda^2_{\max} (|\phi_N\rangle ), 
\end{equation}
where  \(\Lambda_{\max} (|\phi_N\rangle ) =
\max | \langle \chi|\phi_N\rangle |\), with  the maximization being taken over all pure states \(|\chi\rangle\)
that are not genuinely \(N\)-party entangled. 
It was shown in Ref. \cite{qwe10, GGM} that ${\cal E} ( |\phi_N\rangle )$ reduces to
\begin{equation}
\label{label}
{\cal E} (|\phi_N \rangle ) =  1 - \max \{\lambda^2_{{\cal A}: {\cal B}} |  {\cal A} \cup  {\cal B} = 
\{1,2,\ldots, N\},  {\cal A} \cap  {\cal B} = \emptyset\},
\end{equation}
where \(\lambda_{{\cal A}:{\cal B}}\) is  the maximal Schmidt coefficients in the \({\cal A}: {\cal B}\) 
bipartite split  of \(|\phi_N \rangle\).

\section{Quantum dense coding}
\label{sec:QDC}

In this section, we define the capacities of dense coding, when an arbitrary multipartite state is used as a channel, shared between several senders and a single receiver. We first consider the scenario of a noiseless channel and then derive the capacity for a noisy covariant channel. 

\subsection{Quantum dense coding via noiseless channel}
\label{subsec:QDCnoiseless}

The capacity of the quantum dense coding protocol quantifies the amount of classical information that can be transferred via a quantum state used as a channel, when an additional quantum channel, which may be noiseless or noisy, is also available. We begin with the noiseless case. For an arbitrary two-party state, $\rho_{SR}$, shared between the sender, $S$, and the receiver, $R$,  the capacity of dense coding (DC) of $\rho_{SR}$ is given by \cite{qwe7,dcgeneral, dcamader}
\begin{eqnarray}
\label{eq:DC2site}
C (\rho_{SR}) &=& \frac{1}{\log_2 d_{SR}}\max\{\log_2 d_S,\,  \nonumber \\
& &\hspace{4em} \log_2 d_S + S(\rho_R) - S(\rho_{SR})\},
\end{eqnarray}
where \(d_S\) is the dimension of the Hilbert space on which the senders' part of the state $\rho_{SR}$ is defined, and where $d_{SR}$ is that on which the entire state $\rho_{SR}$ is defined. Here, $\rho_R$ is local density matrix  of the receiver's part. The denominator, $\log_2d_{SR}$, is incorporated to make the capacity dimensionless. Note that we are considering the case of unitary encoding (for non-unitary encoding, see \cite{HoroPiani,Shadmansob}), and where the additional quantum channel is noiseless. The noisy case is considered below.  When $ S(\rho_R) - S(\rho_{SR}) >0$, a shared quantum state is better for classical information transmission than any classical protocol with the same resources. It immediately implies that any bipartite pure entangled state is a good quantum channel for dense coding.

In the multipartite regime, let us consider the scenario where there are $N$ senders, $S_{1},\ldots, S_{N}$ and a single receiver, $R$. For the quantum state, $\rho_{S_1\ldots S_{N}R}$, shared between the $N+1$ parties, the capacity of DC with unitary encoding and for noiseless additional quantum channel is given by \cite{qwe7,dcamader}
  \begin{eqnarray}
  \label{eq:DCmult}
C (\rho_{S_1\ldots S_N R}) &=& \frac{1}{\log_2 d_{S_1\ldots S_{N}R}} \max\{\log_2 d_{S_1\ldots S_N}, \nonumber \\
 & & \hspace{0.5em} \log_2 d_{S_1\ldots S_N} + S(\rho_R) - S(\rho_{S_1\ldots S_N R})\},\,\,\,\,\,\,\,\,\,\,\,
\end{eqnarray}
 where $d_{S_1\ldots S_N} = d_{S_1}\ldots d_{S_N}$, with $d_{S_1}$, $\ldots$,  $d_{S_N}$ being the dimensions of  the Hilbert spaces corresponding to the individual senders, and $d_{S_1\ldots S_{N}R}$ is the total dimension of the Hilbert space on which the entire multiparty state is defined. The state is therefore dense codeable if $S(\rho_R) - S(\rho_{S_1\ldots S_N R}) > 0$ . In this paper, whenever we call a multiparty state as dense codeable, it implies that the state is good for dense coding with multiple senders and a single receiver.  
 
 \subsection{Capacity of quantum dense coding through noisy channel}
 \label{subsec:QDCnoise}
 
Consider now the situation of classical information transmission when the additional quantum channel is noisy. We assume that after local unitary encoding, the particles are sent through a noisy quantum channel.  We consider here a particular class of channels, known as the covariant channels, denoted by $\Lambda$. Such a channel is a completely positive map with the property that it commutes with a complete set of orthogonal unitary operators, $\{W_i\}$, i.e.,
\begin{equation}
 \Lambda (W_i \rho W_i^{\dagger}) = W_i  \Lambda (\rho) W_i^{\dagger},  \,\, \forall i.
\end{equation}  
The multipartite dense coding capacity has already been calculated for this  channel \cite{Shadmansob} and is given by
%In this case, the capacity reduces to
  \begin{eqnarray}
\label{Eq:capacity_noisy}
C^{noisy}(\rho_{S_1\ldots S_NR}) &=& \frac{1}{{\log_2d_{S_1\ldots S_N R}}} \max\{\log_2 d_{S_1\ldots S_N}, \nonumber \\
& &\hspace{1em} \log_2 d_{S_1\ldots S_N} + S(\rho_R) - S(\tilde{\rho})\},\,\,\,\,\,\,\,\,\,\,\,
 \end{eqnarray}
where $$\tilde{\rho}=\Lambda \left((U^{min}_{S_1S_2...S_N}\otimes I_R)\rho_{S_1\ldots S_N R} (U^{min\dagger}_{S_1S_2...S_N} \otimes I_R)\right).$$
Here, $U^{min}_{S_1S_2...S_N}$ denotes the unitary operator on the senders' side, which minimizes the von Neumann entropy of $(U_{S_1\ldots S_N}\otimes I_R) \rho_{S_1 \ldots S_NR} (U_{S_1\ldots S_N}^{\dagger}\otimes I_R) $   over the set of unitaries $\{U_{S_1S_2...S_N}\}$, that can be global as well as local. 
It is reasonable from a practical point of view to assume that the senders perform local encoding. Then $U^{min}_{S_1S_2...S_N}$ will have the form given by
\begin{equation}\label{local_unitary_def}
 U^{min}_{S_1S_2...S_N} = U^{min}_{S_1}\otimes U^{min}_{S_2}\otimes...\otimes U^{min}_{S_N}.
\end{equation}
Since only the particles of the senders are sent through the noisy channel (after the unitary encoding), 
%at the sender sides,
the entropy of the receiver's side remains unchanged. 
%This channel is later refereed as correlated channel.    
Depending on the structure of $ \Lambda$, the channel can be either correlated or uncorrelated. 

In Eqs. (\ref{eq:DC2site}), (\ref{eq:DCmult}) and (\ref{Eq:capacity_noisy}), for convenience, we call the second terms within the maximum in the numerators, divided by the denominators, as the corresponding ``raw" capacities. We use the same notation for the raw capacity as the corresponding original capacity, but the context will always make the choice clear.

\subsubsection*{Pauli Channel}
\label{subsec:Paulichannel}

The Pauli channel is an example of a covariant channel.
When  an arbitrary  two-dimensional quantum state is sent through the Pauli channel, the state is rotated by any one of the Pauli matrices, $\sigma_x,\, \sigma_y,\, \sigma_z$, 
%viz.,
%$$
%\sigma_x=\left(
%\begin{array}{cc}
%0 & 1 \\
%1 & 0
%\end{array}\right)
%,\,
%\sigma_y=\left(
%\begin{array}{cc}
%0 & i \\
%-i & 0
%\end{array}\right)
%,\,
%\sigma_z=\left(
%\begin{array}{cc}
%1 & 0 \\
%0 & -1
%\end{array}\right)
%,$$
or left unchanged. If the $\sigma_x$, $\sigma_y$, and $\sigma_z$ act respectively with the probabilities $\lambda_x$, $\lambda_y$, and $\lambda_z$, then the transformed state is
\begin{equation}\label{Eq:De_ex}
\rho'=\lambda_x \sigma_x \rho \sigma_x + \lambda_y \sigma_y \rho \sigma_y + \lambda_z \sigma_z \rho \sigma_z + (1-\lambda_x-\lambda_y-\lambda_z) \rho.
\end{equation}
For $\lambda_x= \lambda_y=\lambda_z=p/3$, the channel represents the depolarizing channel \cite{presskillecture}.

\section{Relationship between multiparty quantum correlations and  dense coding capacity: Noiseless Channel}
\label{sec:relationsnoiselss}

In this section, we establish a generic relation between the capacity of dense coding and various quantum correlation measures, defined in Sec. \ref{sec:QCM&M}. The analytical results obtained in this section are for quantum systems with $(N+1)$-qubits, consisting of $N$ senders and a single receiver. The numerical simulations that are performed to visualize the results are for three-qubit pure as well as mixed states. Throughout this section, we consider the case when the additional quantum channel, that is used post-encoding, is noiseless.

\subsection{Connection between Capacity and Multipartite Quantum Correlations for Pure States}
\label{purestates}
%\label{subsec:relationsnoiseless}

%We consider a $(N+1)$-party state, $\rho_{A_{1}A_{2}...A_{N}B}$, shared between $N$ senders, Alices, and a receiver, Bob, in different locations. The  aim of the sender is to send the classical information in their position to Bob by using quantum dense coding protocol. To achieve this Alice's makes an encoding in their subspace and then send their encoded particle through the noiseless channel to Bob, where Bob apply measurement to retrieve the encoded information. The state $\rho_{A_{1}A_{2}...A_{N}:B}$ is a said to be multiparty entangled state such that the multipartite quantum correlation measure, tangle or discord or generalized geometric measure, is a positive quantity. 

In a bipartite scenario, all the pure states with the same amount of entanglement have equal capacity of dense coding. The entanglement in this case is uniquely classified by the von Neumann entropy of the local density matrices and the capacity is maximal for the maximally entangled states. 

%We now show that this is not the case for multipartite states. Before proceed further, let us identify a multipartite state which can have maximal quantum correlation, with respect to any quantum correlation measure. Then, the multipartite state can act as an unit of a scale for quantum correlations, like maximally entangled state in a bipartite domain. 
We will see that this simple situation is no more true in the multiparty regime. However, it is still possible to obtain a generic relation between capacity and entanglement. 
In a multipartite scenario, quantification of quantum correlations is not unique even for pure states and hence each measure, in principle, identifies its own distinct state  with maximal quantum correlation. Nevertheless, the Greenberger-Horne-Zeilinger (GHZ) state \cite{GHZ} has been found to possess a high amount of multipartite quantum correlation, according to  violation of certain  Bell inequalities \cite{GHZbellineq}, as well as according to several multipartite entanglement measures \cite{qwe10,GM, GGM}. In view of these results,  we compare the properties of an arbitrary $(N+1)$-qubit pure state with that of the  $(N+1)$-qubit generalized GHZ state ($g$GHZ), which is given by
\begin{eqnarray}\label{Eq:gGHZ_def}
 |gGHZ\rangle_{S_{1}S_{2} \ldots S_{N}R}  &=& \sqrt{\alpha} |0_{S_1} \ldots 0_{S_{N}} \rangle |0_R\rangle \nonumber \\ 
 &+&\sqrt{1 - \alpha}e^{i\phi} |1_{S_1} \ldots 1_{S_{N}} \rangle |1_R\rangle,
\end{eqnarray}  
where $\alpha$ is the real number in (0,1) and $\phi \in [0, 2\pi)$. We find that if the capacity of dense coding of an arbitrary $(N+1)$-party state, $|\psi\rangle$, and the $g$GHZ state are the same, then the quantum correlations of these two states may not be the same. However, they follow an ordering, which we establish in the following two theorems. The feature is generic in the sense that it holds for drastically different choices of the quantum correlation measures.
Here on, we skip all the subscripts in the notation of the states, for simplicity.  
%The unequalness of DC capacities between two states having same quantum correlations is generic, i.e. independent of choices of quantum correlation measures.  
 %A generalized GHZ state shared between $(N+1)$-parties is given by
%\begin{eqnarray}
 %|gGHZ\rangle_{A_{1}A_{2} \ldots A_{N}:B} &=& \alpha |0_{A_1} \ldots 0_{A_{N}} \rangle |0_B\rangle \nonumber \\ 
%& & +\beta |1_{A_1} \ldots 1_{A_{N}} \rangle |1_B\rangle, 
%\end{eqnarray}
 %with $|\alpha|^2 + |\beta|^2 =1$.

\noindent {\bf Theorem 1:} \textit{Of all the multiqubit pure states with an arbitrary but fixed multiparty dense coding capacity, the generalized GHZ state has the highest GGM.}

\texttt {Proof.} Scanning over $\alpha$ in Eq. (\ref{Eq:gGHZ_def}), one can obtain an arbitrary value of the GGM. Therefore to prove the theorem, one needs to show that if the multiparty dense coding capacity of an arbitrary $(N+1)$-qubit  pure state  is the same as that of an $(N+1)$-party $g$GHZ state, then the genuine multipartite entanglement, as quantified by the GGM, of that arbitrary pure state is bounded above by that of the $g$GHZ state, i.e.,
\begin{equation}
 {\cal E}(|\psi\rangle) \leq {\cal E}(|gGHZ\rangle).
\end{equation}
%\textit{for any number of qubits}\\

 The multipartite dense coding capacities of the $(N+1)$-party $g$GHZ state and the arbitrary pure state, $|\psi\rangle$, can be obtained by using Eq. (\ref{eq:DCmult}), and  are given respectively by
 \begin{eqnarray}
 \label{boo}
  C(|gGHZ\rangle) 
&=& \frac{N}{N+1} - \frac{\alpha \log_2\alpha + (1-\alpha)\log_2(1-\alpha)}{N+1}\,\,\,\,\,\nonumber
 \end{eqnarray}
and 
 \begin{eqnarray}\label{Eq:capacity_GGHZ}
  C(|\psi\rangle) = \frac{N}{N+1} - \frac{\lambda_R \log_2\lambda_R + (1-\lambda_R)\log_2(1-\lambda_R)}{N+1},\,\,\,\,\,\nonumber
 \end{eqnarray}
where $\lambda_R$ is the maximum eigenvalue of the marginal density matrix,  $\rho_R$, of the receiver's part of the state $|\psi\rangle$.
% and without any loss of generality we assume that $A_{1}A_{2}...A_{N}$ are the senders and $B$ is the receiver. 
The GGMs for the $g$GHZ state and the $\ket{\psi}$ are obtained respectively by
 \begin{equation}
 \label{Ga}
 {\cal E}(|gGHZ\rangle) = 1 - \alpha,
 \end{equation}
% assuming $\alpha \geq 1/2$ while the GGM of  $|\psi\rangle$, is given by
 \begin{equation}
\label{Gal}
 {\cal E}(|\psi\rangle) = 1 - \max[\{{l}_{A}\}],
 \end{equation}
 where we assume that $\alpha \geq \frac{1}{2}$. The set $\{{l}_{A}\}$ contains the maximum eigenvalues of the reduced density matrices of all possible bipartitions of $|\psi\rangle$.  
 Equating the multiparty dense coding capacities for these two states, we obtain
 \begin{equation}\label{alps}
  \alpha = \lambda_R.
 \end{equation} 
 %Since $\rho_R$ is a rank-2 state, one of its eigenvalues, say, 
Note that  $\lambda_R \in \{l_{A}\}$. 
 Let us now consider the two following cases: (1) the maximum in GGM is attained by  $\lambda_R$, and (2) the maximum is attained by an eigenvalue which is  different from $\lambda_R$.\\
\noindent{\bf Case 1:} 
Suppose $\lambda_R = \max [\{l_{A}\}] $.
Then
\begin{eqnarray}
\label{abr}
 {\cal E}(|\psi\rangle) &=& 1- \lambda_R = 1- \alpha 
% {\cal E}(|\psi\rangle) &=& 
={\cal E}(|gGHZ\rangle),
\end{eqnarray}
by using Eq.  (\ref{alps}).

\noindent{\bf Case 2:}
Suppose $\lambda_R \neq \max [\{l_{\cal A}\}]$. Let $ \lambda_R \leq \lambda_0 = \max [\{l_{\cal A}\}]$. Therefore, we obtain 
\begin{eqnarray}\label{daud}
 {\cal E}(|\psi\rangle) &=& 1- \lambda_0  \leq  1- \lambda_R = 1- \alpha \nonumber
% {\cal E}(|\psi\rangle) &\leq &  
={\cal E}(|gGHZ\rangle)
\end{eqnarray}
Hence the proof. \hfill $\blacksquare$

%The above theorem implies that there exists two kinds of pure states in a multipartite domain -- one set  contains pure states having same amount of genuine multipartite entanglement with the gGHZ state, and have equal multipartite capacity of DC as that of the gGHZ state. The other one is  a set of arbitrary  pure states which have same amount of genuine multipartite entanglement as that of the gGHZ state. However, this set of arbitrary $(N+1)$-party states is more advantageous than the gGHZ state for the purpose of multipartite dense coding.

\begin{figure}[t]
\begin{center}
 \includegraphics[width=0.7\columnwidth,keepaspectratio,angle=270]{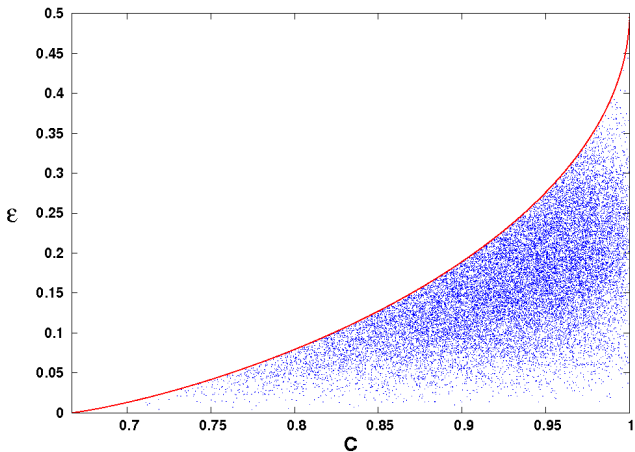}
 \end{center}
\caption{(Color online.) GGM vs. multipartite DC capacity. GGM is plotted as the ordinate while multipartite DC capacity is plotted as the abscissa for  $10^5$ randomly chosen three-qubit pure states,  according to the uniform Haar measure over the corresponding space (blue dots).  The red line represents the generalized GHZ states.  There is a set of states for which, if the capacity matches with a $g$GHZ state, then their GGMs are also equal. For the remaining states, if the capacity is equal to a $g$GHZ state, its GGM is bounded above by that of the $g$GHZ state.  Note that the range of the horizontal axis is considered only when the states are dense codeable. The quantities represented  on both the axes are dimensionless. We are considering the case where the post-encoded states are sent through noiseless channels. }
\label{fig:GGMcapacity}
 \end{figure}

 We randomly generate $10^5$ arbitrary three-qubit pure states by using the uniform  Haar measure on this space and plot the behavior of the GGM versus the DC capacity for these states. As proven in Theorem 1, the scatter diagram populates only a region outside the parabolic curve of the $g$GHZ states. See Fig. \ref{fig:GGMcapacity}. Interestingly, therefore, in the plane of the dense coding capacity and the GGM, there exists a forbidden region which cannot be accessed by any three-qubit pure state. With respect to dense coding in the noiseless case, therefore, the $g$GHZ state is the least useful state among all states having an equal amount of the multiparty entanglement.
 %, as quantified by the GGM.
 %As shown in Theorem 1, as well as in Fig. \ref{fig:GGMcapacity}, the dense coding capacity of any three-qubit pure state is always better than that of the gGHZ, when they both possess same amount of entanglement. In some sense, the gGHZ state, in the noiseless case, is less useful state to share. 
 
We now show that the result is potentially  independent of  the choice of the multipartite quantum correlation measure. Towards this aim, let us now consider the tangle and the discord monogamy score, as multiparty quantum correlation measures. The relations between these two quantum correlation measures and the capacity of DC are established in the following theorem. 

\noindent {\bf Theorem 2:} \textit{Of all multiqubit pure states with an arbitrary but fixed multiparty dense coding capacity, the generalized GHZ state has the highest tangle as well as the highest discord monogamy score. }
%\textit{If the multiparty dense coding capacity of an arbitrary $(N+1)$-party  pure state  is same as the capacity of the $(N+1)$-party gGHZ state, then the tangle as well as discord monogamy score of  arbitrary pure states are bounded above by these quantum correlations of the gGHZ state, provided the receiver is the nodal observer in the monogamy relation.} 

\noindent {\em Note:} The tangle and discord monogamy score are defined here by using the receiver of the DC protocol as the nodal observer. 
%{\bf Is the tangle (for arb. N) independent of the nodal observer?}

\noindent  \texttt{Proof.} The equality of the multipartite dense coding capacities of the $(N+1)$-party $g$GHZ state and the arbitrary pure state, $|\psi\rangle$, implies that $\alpha = \lambda_R \geq 1/2$. Notations are the same as in the proof of Theorem 1.
Note that the tangle  of the $g$GHZ state is $4 \alpha (1 -\alpha)$.
%and $S(\rho_R)$, with the assumption that the nodal observer is the receiver.   
%For the tangle of an arbitrary state with the receiver, R, of the dense coding protocol  as the nodal observer, we have
Therefore, we have 
\begin{eqnarray}
\delta_{{\cal C}^2} (|\psi\rangle) &=& 4 \lambda_R (1 - \lambda_R) - \sum_i {\cal C}^2(\rho_{S_i R})  \nonumber\\
&\leq&  4 \lambda_R (1 - \lambda_R) = 4 \alpha (1 -\alpha) = \delta_{{\cal C}^2} (|gGHZ\rangle).\nonumber\\
\end{eqnarray}
The inequality in the second line is due to the fact that ${\cal C}^2(\rho_{S_iR})$ are non-negative, where $\rho_{S_iR}$ is the density matrix of the sender $S_i$ and the receiver $R$ corresponding to the state $|\psi\rangle$.
Similarly, for the discord score, we have
\begin{eqnarray}
\delta_{\cal D}(|\psi\rangle) &=& S(\rho_R) - \sum_i {\cal D} (\rho_{S_i R})
  \nonumber\\
&\leq& S(\rho_R) = S(\alpha)  = \delta_{\cal D} (|gGHZ\rangle), 
 \end{eqnarray}
since \(0 \leq {\cal D} (\rho_{S_i R})\leq 1\). Here, $S(\alpha)$ 
%denotes the Shannon entropy of $\alpha$, which is 
denotes the von Neumann entropy of the single-side density matrix of the $g$GHZ state.
Hence the proof. \hfill $\blacksquare$

\begin{figure}[t]
\begin{center}
 \includegraphics[width=0.34\columnwidth,keepaspectratio,angle=270]{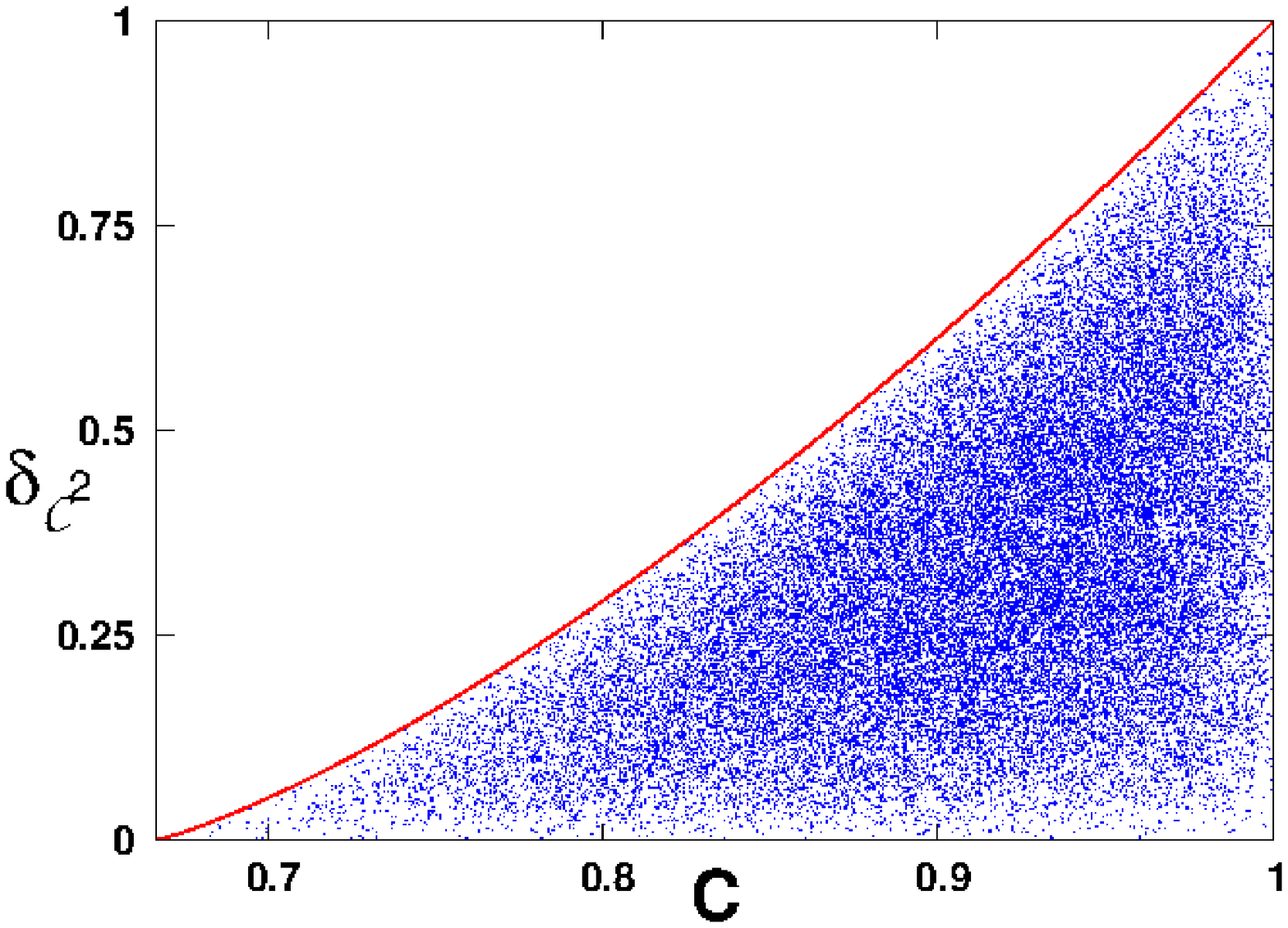} 
 \includegraphics[width=0.34\columnwidth,keepaspectratio,angle=270]{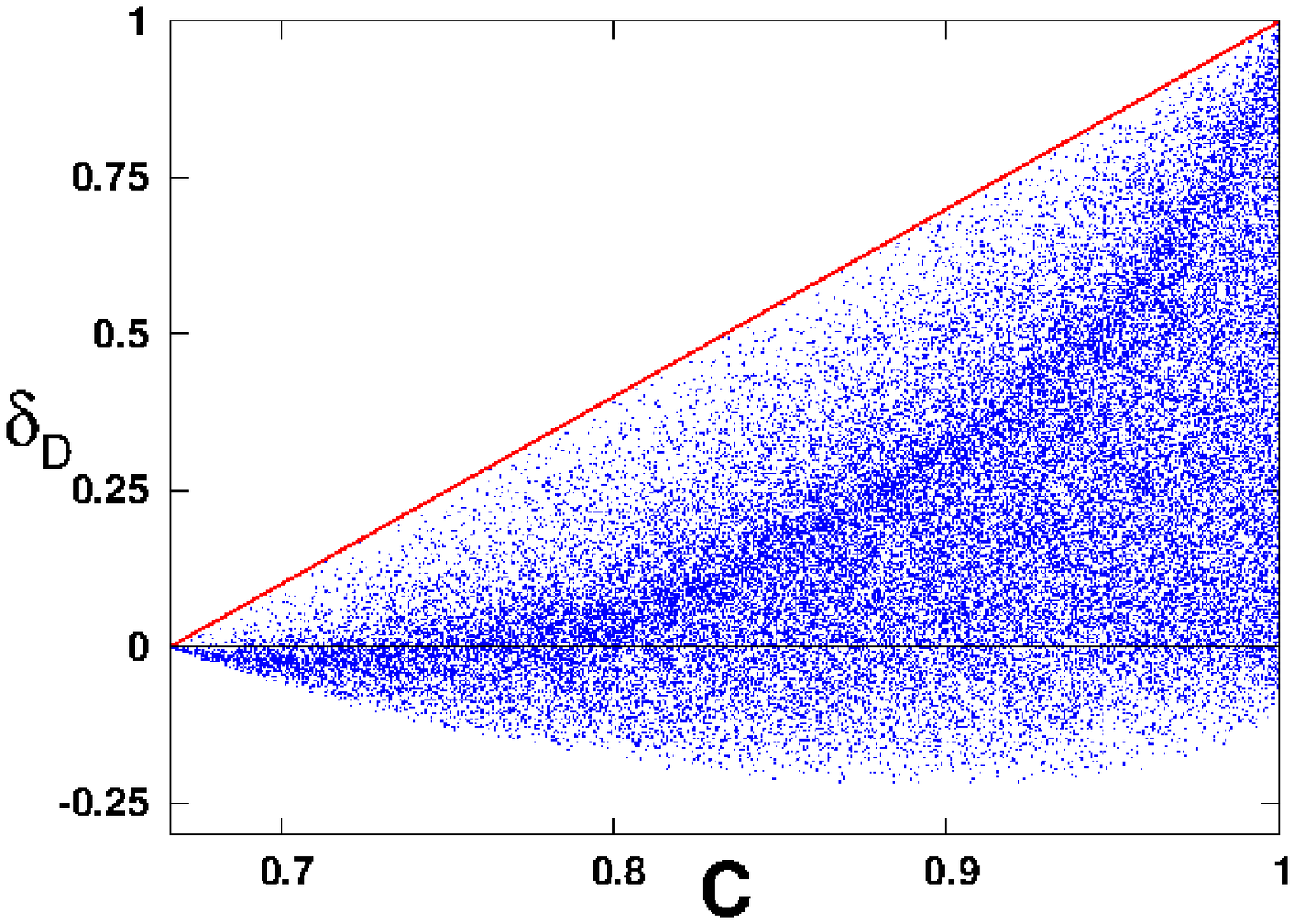}
 \end{center}
\caption{(Color online.) Left: Tangle (vertical axis) vs. multiparty DC capacity (horizontal axis) for randomly generated three-qubit pure states (blue dots).
Right: Discord monogamy score (vertical axis) vs. DC capacity (horizontal axis) for the same states.
In both the cases, the $g$GHZ states give the boundary (red line). The capacity is dimensionless, while the tangle and discord monogamy score are measured in ebits and bits, respectively. All other considerations are the same as in Fig. \ref{fig:GGMcapacity}. }
\label{fig:Tanglecapacity}
 \end{figure}

To visualize the above theorem, we  randomly generate $10^5$ pure three-qubit states, by using the uniform Haar measure in the corresponding space, and  prepare scatter diagrams for tangle versus the multiparty DC capacity (Fig. \ref{fig:Tanglecapacity} (left)) and for the discord monogamy score versus the same capacity (Fig. \ref{fig:Tanglecapacity} (right)). The simulations are clearly in agreement with Theorem 2. In particular, and just like for GGM versus the capacity, the planes of $(C,\delta_{{\cal C}^2})$ and $(C, \delta_D)$ can not be fully accessed by the three-qubit pure states.

\subsection{Capacity vs. Multipartite Quantum Correlations for Shared Mixed States}
\label{mixedstate}

We now investigate the relation between DC capacity and multipartite quantum correlation measures, when the shared state is an arbitrary $(N+1)$-party mixed state. In this case, to establish such connection, the main difficulty  is that there are only a few quantum correlation measures available  which can be computed. In this case, therefore, we consider the discord monogamy score as the multipartite quantum correlation measure, since quantum discord can be numerically calculated for arbitrary bipartite systems, and investigate its connection with the DC capacity. 

\begin{figure}[t]
\begin{center}
 \includegraphics[width=0.7\columnwidth,keepaspectratio,angle=270]{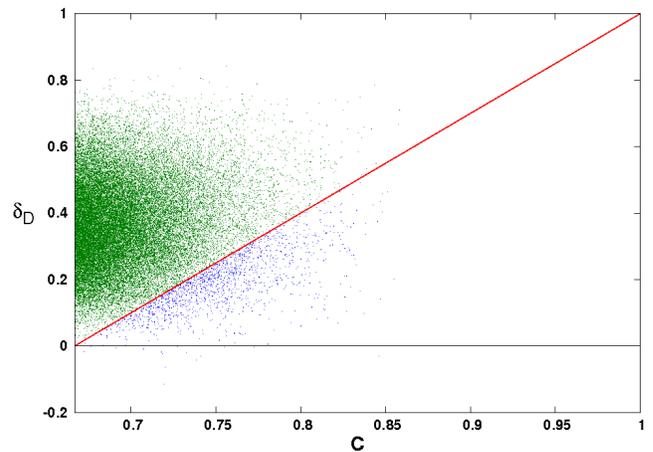}
 \end{center}
\caption{(Color online.) Discord monogamy score vs. multipartite DC capacity for Haar uniformly generated rank-2 three-qubit states. The red line represents the $g$GHZ states. About 1.85\% of the randomly generated states lie below the red line, and are represented by blue dots. The remaining, represented by green dots, lie above the red line. The horizontal axis is dimensionless while the vertical one is measured in bits.}
\label{fig:Discordcapacitymixed}
 \end{figure}

In Fig. \ref{fig:Discordcapacitymixed}, we randomly generate $10^5$ mixed states of rank-2 in the space of three-qubit  states and plot the discord monogamy score with respect to the  DC capacity. The random generation is with respect to the uniform Haar measure induced from that in the appropriate higher-dimensional pure state space. The numerical simulation reveals that Theorem 2 does not hold  for rank-2  (mixed) three-qubit states. In particular, we find that if a $g$GHZ state and a rank-2 three-qubit mixed state have the same discord monogamy score, then sharing the $g$GHZ state is usually more beneficial than the mixed state, for performing the multiparty DC protocol. More precisely, among randomly generated $10^5$ rank-2 states, there are only  $1.85\%$ states  which satisfy Theorem 2. We will later show that a similar picture is true for the noisy channel. This implies that in the presence  of noisy environments, irrespective of whether the noise is afflicted before or after encoding, it is typically better to share a $g$GHZ state among states with a given discord monogamy score, from the perspective of DC capacity. Before presenting the results obtained by using numerical simulations for higher-rank mixed states, let us  discuss the behavior of the DC capacity, as enunciated in the following proposition. We will find that it can be used to intuitively understand the numerical results for higher-rank states presented below.

\noindent{\bf Proposition 1:} \textit{An arbitrary $(N+1)$-qubit (pure or mixed) state is dense codeable if the maximum eigenvalue of the $(N+1)$-party state is strictly greater than the maximum eigenvalue of its reduced state at the receiver's side.}
%\begin{equation}\label{majo_cond}
%\sum \lambda_{S_1S_2...S_N R}^{\max} > \lambda_R^{\max}
%\end{equation}
% \textit{where the summation is taken over all eigenvalues other than the maximum one.}

\noindent\texttt{ Proof:}
An $(N+1)$-qubit (pure or mixed) state, $\rho_{S_1S_2 \ldots S_NR}$, is multiparty dense codeable with $N$ senders, $S_1,S_2, \ldots, S_N$, and a single receiver, R, if and only if the von Neumann entropy of the reduced state at the receiver's side is greater than that of the state $\rho_{S_1S_2 \ldots S_NR}$, i.e.,
\begin{equation}
\label{densecondition}
 S(\rho_{R}) > S(\rho_{S_1S_2 \ldots S_N R}).
\end{equation}
%If the state $\rho_{R}$ is a rank-2 state, then the corresponding 
Let the eigenvalues, in descending order, of the state $\rho_R$, be given by
$\lambda_R = \{\lambda^1 \geq \frac{1}{2}, 1 - \lambda^1\}$.
Let the eigenvalues of the state $\rho_{S_1S_2 \ldots S_N R}$ be $ \lambda_{S_1S_2 \ldots S_N R} =\{\mu^i\}_{i=1}^r $, where $r$ is the rank of the matrix, and where the $\mu^i$'s are arranged in descending order. Specifically, $\mu^1$ gives the largest eigenvalue of 
 $\rho_{S_1S_2 \ldots S_N R}$.
Now, the ordering between the highest eigenvalues of $\rho_{R}$ and $\rho_{S_1S_2 \ldots S_N R}$, i.e., between $\lambda^1$ and $\mu^1$, can have three possibilities, i.e.,
%\begin{eqnarray}
$\lambda^1 > \mu^1$, or they are equal, or $\lambda^1 < \mu^1$.
%\end{eqnarray}

 Let us assume that $\lambda^1 \geq \mu^1$. Then, invoking the condition of majorization \cite{RajendraBhatia}, we have 
 \begin{equation*}
\lambda_R \succ \lambda_{S_1S_2...S_N R},
 \end{equation*}
which implies 
 \begin{equation}
  S(\rho_{R}) \leq S(\rho_{S_1S_2...S_N R}).
 \end{equation}
%The above inequality is due to the relation between majorization and von Neumann entropy. 
It immediately implies that the state is not dense codeable. Therefore, to obtain dense codeability of  $\rho_{S_1S_2...S_N R}$, we must have $\lambda^1 < \mu^1$.\\
Hence the proof. \hfill $\blacksquare$

 Although the above proposition has been presented for qubit systems, it is also valid for an arbitrary (pure or mixed) $(N+1)$-party quantum state in arbitrary dimensions, provided $\rho_R$ is of rank 2.
 
\begin{figure}[t]
\begin{center}
 \includegraphics[width=0.8\columnwidth,keepaspectratio, angle = 270]{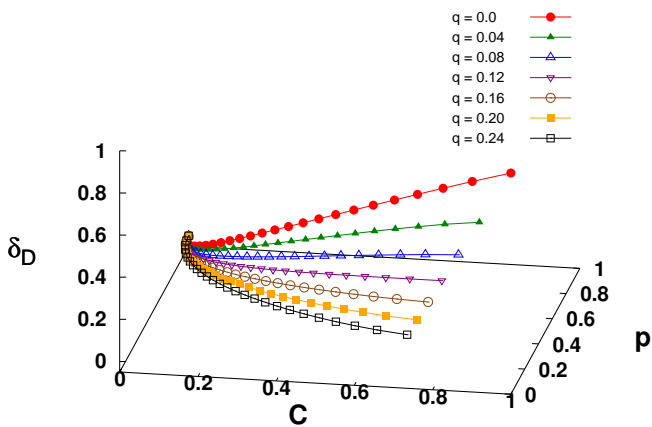}
 \end{center}
\caption{(Color online.) Discord monogamy score and the raw  DC capacity are plotted against the mixing parameter $p$, for the rank-8 state, $\rho_8 = (1-p)\rho + \frac{p}{8}I$. Here $\rho=q|GHZ\rangle\langle GHZ|+(1-q)|GHZ'\rangle\langle GHZ'|$. Each value of $q$ provides a curve, and we present several exemplary curves in the figure. 
%The surface is obtained from various choices of $q$. The state is dense codeable, when $C>0.667$. 
All the quantities plotted are dimensionless, except $\delta_D$, which is measured in bits. }
\label{fig:Densefull}
 \end{figure}

Let us now move to mixed states with higher-rank. Numerically,  to obtain high-rank three-qubit mixed states, one possibility is to generate pure states with more than three parties. For example, to obtain arbitrary rank-4 states of three qubits,  5-qubit pure states can be created randomly, and then  two parties traced out. However, numerical searches become inefficient with the increase of number of parties \cite{HaydenWinter}. To overcome this problem,   we create  mixed states, $\rho_8$, of full rank, given by 
\begin{equation}
 \rho_8 = (1-p)\rho + \frac{p}{8} I_8,
\end{equation}
by choosing $\rho$  as  arbitrary rank-2 three qubit states, generated randomly from the three-qubit pure states, and where $I_8$ is the identity matrix on the three-qubit Hilbert space.  Moreover, we consider those set of states, $\rho$, which are dense codeable.   In that case, we find that  its DC capacity remains nonclassical only for very small values of the mixing parameter $p$. In  Fig. \ref{fig:Densefull}, we specifically consider the  full rank state, $\rho_8$, with $\rho$ given by 
\begin{eqnarray}
\rho=q|GHZ\rangle\langle GHZ|+(1-q)|GHZ'\rangle\langle GHZ'|,
\end{eqnarray}
where $|GHZ'\rangle=\frac{1}{\sqrt{2}}(|000\rangle-|111\rangle)$.  We now plot, in Fig. \ref{fig:Densefull}, the discord monogamy score and the raw DC capacity with respect to the mixing parameter $p$.  For $q = 1$ or $q=0$, and $p = 0$, the capacity is maximum and $\delta_D$ also gives a maximum. Fig. \ref{fig:Densefull} shows that there is a small region in which the state remains dense codeable, only when $\delta_D$ is very high. It is plausible that the capacity of dense coding for mixed states decreases with the increase of rank of the state.  This is intuitively understandable from the condition in Proposition 1, since the typical high-rank state can have eigenvalues more distributed than the typical low-rank state. Therefore, the maximal eigenvalue of a shared state typically gives a lower value than that of the receiver's side, and the condition in Proposition 1 is thereby satisfied for a very small set of states. 
 
%  Here we see as the rank increases to 8 (corresponding to non zero p) the state is still dense codeable provided that Eigenvalues of $\rho_8$ majorizes the eigenvalues of $\rho_{8_A}$. As p; increases Eq(\ref{majo_cond}) is violated, and $\rho_8$ can no longer be useful for coding.

\section{Relation between Capacity of Dense coding  and Quantum Correlations for Noisy Channels}
\label{sec:relationsnoisy}

In this section, we consider the DC capacity of the noisy channels, for both correlated and uncorrelated noise,  for the $(N+1)$-party state, $\rho_{S_1S_2 \ldots S_N R}$, shared between $N$ senders and a single receiver. Here we assume that the $N$ senders individually apply local unitary operations on their parts of the shared state and send their encoded parts through a covariant noisy channel (see Sec. \ref{subsec:QDCnoise}). We now address the extent to which  the relation, established in Sec. \ref{sec:relationsnoiselss}, between multiparty DC capacity for the noiseless channel and multipartite quantum correlation measures, in the case of pure shared quantum states, still remains valid for the noisy channel scenario. To this end, we consider two extreme scenarios, one in which the noise between the different sender qubits are fully correlated, and another in which the same are uncorrelated.  

\subsection{Fully correlated Pauli channel}
\label{subsec:fullycorrelatedPauli}
 %A Pauli channel acting on a (N+1) party mixed state is given in Eq. (\ref{PCmulti}). It is said to be a fully correlated Pauli channel if same $W_{mn}$ acting on all the senders, i,e, $q_{\{m_in_i\}} = q_{mn}$ for all channels i. 
 
 %For a qubit system, the mutually orthogonal complete set of unitary states $W_{mn}$ defined in Eq. (\ref{Ortho_uni_def}) are the Pauli matrices discussed in Sec. \ref{subsec:Paulichannel}.
 %Now denoting $\sigma_x = \sigma^1, \sigma_y = \sigma^2, \sigma_z = \sigma^3, \mathbb{1} = \sigma^0$.
 An $(N+1)$-qubit state, $\rho_{S_1S_2 \ldots S_N R}$, after being acted on by the fully correlated Pauli channel, is given by
 \begin{eqnarray}
\label{Pauli_channel_def}
 &\Lambda^{P}(\rho_{S_1S_2 \ldots S_N R}) = \sum_{i=0}^3 q_m(\sigma^m_{S_1}\otimes \cdots \otimes\sigma^m_{S_N}\otimes I_R) \nonumber \\
 &\rho_{S_1S_2 \ldots S_N R}(\sigma^m_{S_1}\otimes \cdots \otimes\sigma^m_{S_N}\otimes I_R),
\end{eqnarray}
 where $\sum_{m=0}^3q_m = 1$, and $q_m \geq 0$, and where we denote, for simplicity, $\sigma_x = \sigma^1, \sigma_y = \sigma^2, \sigma_z = \sigma^3,$ and the identity matrix as $\sigma^0$ for the sender qubits. The receiver qubit is acted on only by the identity operator, which we denote by $I_R$. 
 
 We now establish the parallel of the ordering in Theorem 1 for the fully correlated Pauli channel.   
 %
%Let us now consider only a set of three-qubit states for which the GGM as well as the capacity of DC, in case of the noiseless channel, are same with that of the gGHZ state.
% in Sec. \ref{sec:relationsnoiseless}).
% states that the GGM of an arbitrary 3-qubit pure state is always less than or equal to the GGM of the gGHZ state, when dense coding capacity for both the states 
% We now show that the presence of noise can introduce ordering between these set of states and the gGHZ state. 
%the reverse of the theorem holds.
%In the following theorem, we establish the ordering between that set of arbitrary three-qubit states and the gGHZ states.

\noindent{\bf Theorem 3:} \textit{If the multiparty dense coding capacity of an arbitrary three-qubit pure state, $|\psi\rangle$,  is the same as that of the $g$GHZ state in the presence of the fully correlated Pauli channel, then the genuine multipartite entanglement, GGM, of that arbitrary pure state is bounded below by that of the $g$GHZ state, i.e.,} 
\begin{equation}
 {\cal E}(|\psi\rangle) \geq {\cal E}(|gGHZ\rangle),
\label{eq:opposite}
\end{equation}
\textit{provided the following two conditions hold: (i) the largest eigenvalue of the noisy $|\psi\rangle$  state  is
%, $\lambda_1$, of $\Lambda^{P}((U^{min}_{S_1S_2}\otimes\mathbb{1}_R)\rho_{S_1S_2R}(U^{min\dagger}_{S_1S_2}\otimes\mathbb{1}_R))$ is 
bounded above by} $max\{q_1+q_2, 1-q_1-q_2\}$, \textit{and (ii) the receiver's side gives the maximum eigenvalue for the GGM of $|\psi\rangle$.}

\noindent \texttt{Proof:}
The capacities of multiparty dense coding of the $g$GHZ state and the three-qubit pure state, $|\psi\rangle$, after being acted on by the  correlated noisy channel, can be obtained from Eq. (\ref{Eq:capacity_noisy}), and are given respectively by
\begin{equation}\label{Eq:gGHZ_noi_capa}
 C^{noisy}_c (|gGHZ\rangle) = \frac{2}{3} + \frac{H(\alpha) - S(\tilde{\rho}_{gGHZ})}{3}
\end{equation}
and
\begin{equation}\label{Eq:arbitrary_noi_capa}
 C^{noisy}_c (|\psi\rangle) = \frac{2}{3} + \frac{H(\lambda_R) - S(\tilde{\rho}_{\psi})}{3}
\end{equation}
where $\tilde{\rho}_{gGHZ} = \Lambda^{P}((U^{min}_{S_1S_2}\otimes I_{R})|gGHZ\rangle \langle gGHZ|$ $(U^{min\dagger}_{S_1S_2}\otimes I_{R}))$ with $U^{min}_{S_1S_2}$ being the unitary operator at the senders' part that minimizes the relevant von Neumann entropy  (see Sec. \ref{subsec:QDCnoise}). Here, we are considering only those cases for which the (noisy) capacities of both the $g$GHZ state as well as of the $|\psi\rangle$ are non-classical, i.e., the corresponding noisy states are dense codeable. Replacing  \(|gGHZ\rangle\) by \(|\psi\rangle\) in \( \tilde{\rho}_{gGHZ}\), one obtains \( \tilde{\rho}_{\psi}\). The $U_{S_1S_2}^{min}$ is of course a function of the input state. Here , $\lambda_R \geq \frac{1}{2}$ denotes the maximum eigenvalue of the reduced density matrix $\rho_R$ of $|\psi\rangle$. For $ 0\leq x \leq 1$, $H(x)$ denotes the binary entropy function $-x\log_2 x -(1-x)\log_2 (1-x)$.
%If $A_1$,$A_2$ are well separated from each other, then they can only perform local unitary operations so w
%Moreover, we assume that $U^{min}_{S_1S_2} = U^{min}_{S_1}\otimes U^{min}_{S_2}$. 
For the $g$GHZ state, 
%we obtain that any local unitary operators 
%$U^{min}_{S_1S_2} = \mathbb{1}_{S_1}\otimes\mathbb{1}_{S_2}$ and hence 
the von Neumann entropy of the resulting state after sending through the fully correlated Pauli channel is  $S(\tilde{\rho}_{gGHZ}) = H(q_1 + q_2)$, which is independent of the choice of the local unitary operators. 

Equating Eqs. (\ref{Eq:gGHZ_noi_capa}) and (\ref{Eq:arbitrary_noi_capa}), we have,
\begin{eqnarray}\label{Eq:prior_proof}
 H(\alpha) &=& H(\lambda_R) + [H(q_1 + q_2) - S(\tilde{\rho}_{\psi})]\nonumber \\
	   &=& H(\lambda_R) + [H(q_1 + q_2) - H(\{\lambda_i\})],
\end{eqnarray}
where $\{\lambda_i\}_{i=1}^{8}$ are the eigenvalues of $\tilde{\rho}_{\psi}$ in  descending order. Here $H(\{\lambda_i\}) = -\sum_i \lambda_i \log_2 \lambda_i$.
If we assume that $\lambda_1 \leq \max\{q_1+q_2, 1- q_1-q_2\}$, we have $\{\lambda_i\} \prec \{q_1+q_2, 1- q_1-q_2\}$. The relation between  majorization and  Shannon entropy \cite{RajendraBhatia} then implies that $ H(q_1 + q_2) \leq H(\{\lambda_i\})$. Therefore, from Eq. (\ref{Eq:prior_proof}), we have
\begin{eqnarray}
\label{Eq:noisyalphlambda}
 H(\alpha) \leq H(\lambda_R) \Rightarrow \alpha \geq  \lambda_R,
\end{eqnarray}
where we assume $\alpha \geq \frac{1}{2}$.

The GGM for the $g$GHZ state and the three-qubit state, $|\psi\rangle$, are respectively given by
%\begin{equation}
$ {\cal E}(|gGHZ\rangle) = 1-\alpha$ %\end{equation}
and %\begin{equation} 
${\cal E}(|\psi\rangle) = 1 - \lambda_{max} $,
%\end{equation}
where $\lambda_{max}$ is the maximum eigenvalue among the eigenvalues of all the local density matrices of $|\psi\rangle$. 
If we assume that the eigenvalue from the receiver's side attains the maximum, i.e., if  $\lambda_R = \lambda_{max}$,  using Eq. (\ref{Eq:noisyalphlambda}), we obtain
\begin{equation}
 {\cal E}(|gGHZ\rangle) = 1-\alpha \leq 1-\lambda_R = {\cal E}(|\psi\rangle).
\end{equation}
Hence the proof. \hfill $\blacksquare$

%Note here that the above theorem does not violate the continuity, when one approaches from the noisy to the noiseless limit. This is due 
%Note here that when the noise parameter vanishes, 
%to the fact that these are the states  have same amount of genuine multipartite 
%entanglement and DC capacity as the gGHZ state in the noiseless case. 
%The above theorem shows that  due to noise, those states to have similar dense coding capability as the gGHZ state require more genuine multipartite entanglement. In some sense,
%In other words, t
The above theorem ekes out a subset of the pure three-qubit state space, for which the $g$GHZ state is more robust with respect to multiparty DC capacity, against fully correlated Pauli noise, as compared to any member of the said subset, provided the $g$GHZ and the said member have equal multiparty entanglement, as quantified by their GGMs. This specific subset of states are those which satisfy both the conditions $(i)$ and $(ii)$. The situation, at least for this specific subset, has therefore exactly reversed with respect to the noiseless scenario, as enunciated in Theorem 1. For a given amount of multiparty entanglement content, as quantified by the GGM, the $g$GHZ state can now be better than other pure states, with respect to the multiport classical capacity. The noisy quantum channel can therefore reverse the relative capabilities of classical information transfer of different states in multiparty quantum systems. The result is much more general than what is contained in Theorem 3. First of all, the result in Theorem 3 holds even if we replace the GGM as the multiparty quantum correlation measure by the tangle or the discord monogamy score, provided we consider the set of three-qubit pure states for which the two-party concurrences or quantum discords respectively vanish, and the receiver is used as the nodal observer. Comparing now with Theorem 2, we see that the phenomenon of the inversion of the relative capabilities for classical information transfer is generic in this sense: it applies irrespective of whether the GGM, or the tangle, or the discord monogamy score is used to measure the multiparty quantum correlation content. Secondly, we will show below that the phenomenon of reversal of information carrying capacity with the addition of noise actually holds for a much larger class of states than the ones covered by the conditions $(i)$ and $(ii)$ in Theorem 3. We resort to numerical searches by generating Haar uniform three-qubit pure states for this purpose. The following picture is therefore emerging. Given a three-qubit pure state, $|\psi\rangle$, and a $g$GHZ state with the same multiparty quantum correlation content, the multiparty DC capacity of the $g$GHZ state is much less affected by noise than a large class of $|\psi\rangle$, and in many cases, the ordering of the capacities can get reversed in the noisy case as compared to the order in the noiseless case. 

 % since their dense coding capacity decreases much more than that of the gGHZ state. 
\begin{figure}[t]
\begin{center}
 \includegraphics[width=0.7\columnwidth,keepaspectratio,angle=270]{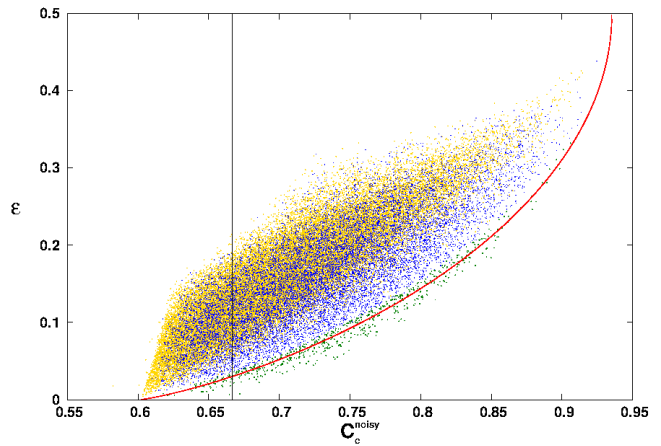}
 \end{center}
\caption{(Color.)   GGM (vertical axis)  vs. the raw DC capacity (horizontal axis) under the fully correlated Pauli channel, when the shared state is an arbitrary three-qubit pure state (orange, green, and blue dots) or the $g$GHZ state (red line). For all the states, the noise in the channel is fixed to 
%$H(\{q_1+q_2,1-q_1-q_2\}) = 0.19$.
0.19. We choose  $q_0 = q_3 = 0.485$, and $q_1 = q_2 = 0.015$ as noise parameters for the arbitrary as well as the $g$GHZ state. This corresponds to the Case 1 of Sec. \ref{subsec:fullycorrelatedPauli}. See the discussion there for further details. Both the axes are dimensionless. The vertical line at the $C^{noisy}_c=2/3$ helps to readily read out the actual capacity from the raw capacity.} \vspace{-1em}
%In the presence of moderate noise, all the arbitrary three-qubit pure states are shifted above the $g$GHZ curve. 
%Orange dots are those states which do not satisfy condition $(ii)$ of Theorem 3 while blue dots are the ones which satisfy both the conditions. The GGM of the arbitrary states (orange and blue dots) is bounded below by the $g$GHZ state  for a fixed value of DC capacity.  The points left to the vertical black line are not useful for dense coding protocol, which is more in number than in Fig. \ref{fig:GGMcapacitynoise}. Both the axes are dimensionless.}
\label{fig:GGMcapacitynoiseMax}
 \end{figure}

%Let us now 
%We now check numerically whether the two constraints stated in the above theorem are satisfied by the three-qubit states which are generated randomly by using Haar measure, i.e.,  w
%We numerically check that the effects of noise on the three-qubit pure states which satisfy the conditions:(i) $\lambda_1 \leq \max\{q_1+q_2, 1-q_1-q_2\}$, and (ii) the maximum eigenvalue in GGM is obtained from the receiver's side, $\rho_R$, i,e, $\lambda_R$ is maximum among all the eigenvalues of the single site density matrices of the three-qubit state.
 To perform the numerical searches, we first  observe that the  $C^{noisy}_c(|gGHZ\rangle)$ depends on the sum of the two parameters $q_1$ and $q_2$ (or $q_0+q_3$). By fixing $q_1+q_2 = c$ (or $q_0+q_3 = 1-c$), one can set the noise parameter for the $g$GHZ state. However, the situation for an arbitrary state, $\ket{\psi}$, is more involved, for which the capacity of dense coding, $C^{noisy}_c(|\psi\rangle)$, depends individually on all the $\{q_i\}$.
 %To visualize the above theorem and conditions (i) and (ii), the $\{q_i\}$ should have some fixed values and 
 %Note that $C^{noisy}_c(|gGHZ\rangle)$ depends on the sum of two parameters of the noisy channel i.e., $q_1 + q_2$. Intuitively, one expects that if the weights of $\{q_1, q_2\}$ are almost equal, i.e., $ H(\{q_1 + q_2,1 - q_1 - q_2\}) \backsimeq 1$, the gGHZ state does not remain dense codeable. This is because that all the $\sigma_i$'s rotate the state by same amount and changes it to a maximally mixed state. Hence it is interesting to consider,  when  $ H(\{q_1 + q_2,1 - q_1 - q_2\})$ is small in which case the gGHZ state can keep its quantumness and is still advantageous for dense coding protocol.
%In case of arbitrary state the value of $C^{noisy}_c(|\psi\rangle)$ will depend on the structure of $\{q_i\}$, where the choices of $\{q_i\}$ are arbitrary. To validate the condition (i) and (ii), it is required to fix the values of $\{q_i\}$.
 To quantify the randomness of $\{q_i\}$, and indeed the noise in the channel, we consider the Shannon entropy, $H(\{q_i\})$. 
 %star12 Here, we are considering only those cases for which the noisy capacities of both the gGHZ state as well as of the $|\psi\rangle$ are non-classical, i.e., the corresponding noisy states are dense codeable. 
We now consider two extreme cases: one for which $H(\{q_i\})$ is maximum and the other in which the same is a minimum, both subject to the constraint $q_1+q_2 = c$, where $0\leq c \leq 1$.
 The maximum of $H(\{q_i\})$ is attained when $q_1=q_2=c/2$ and $q_0=q_3=(1-c)/2$, while the minimum is obtained when any one of  the $q_1$ and $q_2$ and any one of  $q_0$ and $q_3$ are zero.
 % and similarly between $q_0$ and $q_3$ one of them is set to zero. 
 It is also evident from Eq. (\ref{Eq:gGHZ_noi_capa}), that one should deal with a very low or very high values of $c$, for the state to remain dense  codeable.
 %The dense codeable arbitrary state is obtained for low value of $c$ (or $1-c$) and exploit over the freedom over choosing $\{q_i\}$.

\begin{figure}[t]
\begin{center}
 \includegraphics[width=0.6\columnwidth,keepaspectratio,angle=270]{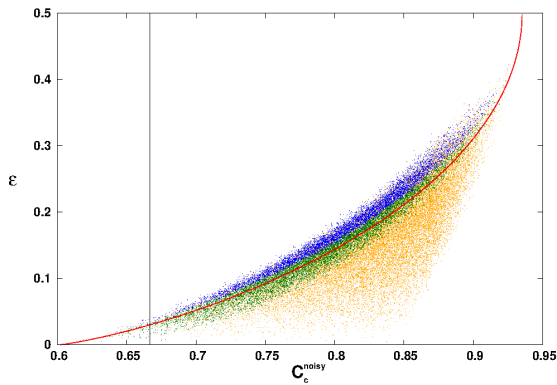}
 \end{center}
\caption{(Color.) GGM (vertical axis)  vs. DC capacity (horizontal axis) under the fully correlated Pauli channel. In this case, we choose $\{q_i\}$ as $q_0 = 0.93, q_1 = 0.01, q_2 =  0.02, q_3= 0.04$. This corresponds to the Case 2 of Sec. \ref{subsec:fullycorrelatedPauli}.  We  randomly (Haar uniformly) generate $5\times 10^4$ three-qubit pure states. See text for further details. Both axes represent dimensionless quantities. The vertical line at $C^{noisy}_c=2/3$ has the same function as in Fig. \ref{fig:GGMcapacitynoiseMax}.} \vspace{-1em}
%Among them,  the maximum in GGM from the receiver's side is obtained for the $33\%$ of states (blue + green dots), satisfying condition $(ii)$ of Theorem 3. Green points are those points which do not satisfy condition $(i)$ while orange are the ones which neither satisfy condition $(i)$ nor $(ii)$. Blue points represent  states which satisfy both the conditions. Red line is for the $g$GHZ state. Most of the states (orange and green points), violate Theorem 3 while blue points satisfy Theorem 3.  Both the quantities plotted in the horizontal  and the vertical axes are dimensionless.   }
\label{fig:GGMcapacitynoise}
 \end{figure}

We now randomly generate  $5 \times 10^4$ three-qubit pure states with a uniform Haar measure over that space, and investigate the two extreme cases mentioned above, for fixed
% to a very low value, such that 
$H(q_1+q_2) = 0.19$.  We choose the two sets of values for the $q_i$'s as follows -- Case 1: $q_0 = q_3 = 0.485$, and $q_1 = q_2 = 0.015$ (see Fig. \ref{fig:GGMcapacitynoiseMax}), and Case 2:  $q_0 = 0.93, q_1 = 0.01, q_2 =  0.02, q_3= 0.04$ (see Fig. \ref{fig:GGMcapacitynoise}). For fixed $H(q_1+q_2) = 0.19$, Case 1 is an example for high noise, and corresponds to the case when $H(\{q_i\})$ is a maximum subject to the constraint $H(q_1+q_2) = 0.19$, which is the same as the constraint $q_1 + q_2 = 0.03$. Case 2 is an example of low noise, and corresponds to a situation that is close to the case when $H(\{q_i\})$ is a minimum subject to the constraint $H(q_1 + q_2) = 0.19$. We present the low noise case, when the configuration is slightly away from the analytical minimum to provide a more non-trivial example. 
%present in the system while the second one is the case of low noise.

\underline{Case 1} (Fig. \ref{fig:GGMcapacitynoiseMax}): In presence of high noise, we observe that almost all the randomly generated  states have shifted to above  the $g$GHZ state (red line) in the plane of GGM and the raw capacity, $C_c^{noisy}$. As expected, one-third of the randomly generated states satisfy condition $(ii)$ of Theorem 3. A significantly large fraction (98.6\%) of them further satisfies condition $(i)$. They are represented by  blue dots in Fig. \ref{fig:GGMcapacitynoiseMax} and lie above the $g$GHZ line. The remaining 1.4\% are represented by green dots, and may lie below or above the $g$GHZ curve. The further states are represented by orange dots.  %All points lie above the red line that corresponds to the $g$GHZ states. 
Note that we have plotted the raw capacity in Fig. \ref{fig:GGMcapacitynoiseMax}.
% represent the $33\%$ of the three-qubit arbitrary states which satisfy the condition (ii) as well as condition (i), while the orange dots represent states  which violate both the conditions. 

\underbar{Case 2} (Fig. \ref{fig:GGMcapacitynoise}): For low noise, the randomly generated states may fall below or above the red line of the $g$GHZ states. Again, one-third of the generated states satisfy condition $(ii)$. $45.6\%$ of them satisfy condition $(i)$, are represented by blue dots, and fall above the red line. The remaining $54.4\%$ of them are represented by green dots, and can be below or above the $g$GHZ line. The other two-thirds are represented by orange dots, and can again be either below or above the $g$GHZ line.

The occurrence of the randomly generated states both below and above the curve for the $g$GHZ states on the plane of the GGM and the capacity is expected from continuity arguments, for low noise.
%of $\{q_i\}$, the condition (ii) is satisfied by $33\%$ of the arbitrary three-qubit pure states (both blue and green dots), and among them, $68\%$ of states  satisfy the first condition (i) (depicted by the blue points). The orange  points represent the states which defy both the conditions. In this case,  all the arbitrary three qubit states are distributed on either side of the gGHZ state (depicted in red line) which is not seen in Case 1. This is  due to  continuity that the states, in the presence of low noise, behaves similarly like the case without noise.  
However, if one makes a comparison between Figs. \ref{fig:GGMcapacity} and \ref{fig:GGMcapacitynoiseMax}, it is revealed that arbitrary three-qubit pure states require higher amount of multipartite entanglement than the $g$GHZ states to keep themselves dense codeable in the presence of moderate noise.

%Interestingly, Theorem 3 also holds for tangle and discord monogamy score if one considers a set of three-qubit pure states for which the square of concurrences or discords of all the reduced density matrices of the shared states vanish, and the receiver is acted as the nodal observer. Numerically, we also find that like GGM, the DC capacity behaves in a similar way with the tangle or discord monogamy score under fully correlated Pauli channel. 
We have also numerically analyzed  the randomly generated states by replacing the GGM with the tangle and with the discord monogamy score. We find the behavior of the DC capacity with these multiparty quantum correlation measures to be similar to that between the DC capacity and the GGM. However, the GGM is more sensitive to noise than tangle or discord monogamy score, in the sense that in the presence of small values of  noise parameters, the percentages of states which are below the $g$GHZ state is much higher in the case of the monogamy score measures than for the GGM. 

Therefore, Theorem 3 and  the numerical simulations strongly suggest that in the presence of fully correlated Pauli noise, the ratio of multipartite entanglement to the DC capacity of the $g$GHZ state increases at a slower rate than that of the arbitrary three-qubit pure states, irrespective of the choice of the multiparty quantum correlation measure. 

\begin{figure}[t]
\begin{center}
  \includegraphics[width=0.6\columnwidth,keepaspectratio,angle=270]{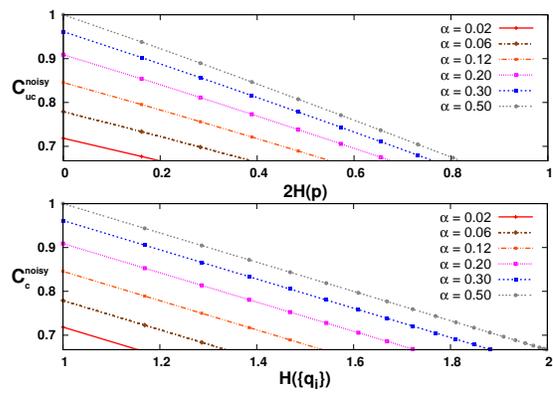}
 \end{center}
\caption{(Color online.) DC capacity vs. noise for various choices of $\alpha$ in the $g$GHZ state. In the top panel, the capacity of DC is plotted against the noise  of the depolarizing channel, while in the  bottom one,  the DC capacity is plotted with respect to the noise in the fully correlated Pauli channel, 
%with the sum of the noise parameter $q_1 + q_2$ in the fully correlated Pauli channel, 
for the $g$GHZ state. Different curves correspond to different values of $\alpha$. The vertical axis starts from 2/3, below which the states are not dense codeable.  The states remain dense codeable in the presence of moderate to high Pauli noise while this is not the case for the uncorrelated depolarizing channel. The horizontal axes are measured in bits. All other quantities are dimensionless.}
\label{fig:corr_vs_Depolarising}
 \end{figure}
 
  \vspace{-1em}
 \subsection{Uncorrelated Pauli channel}
 \label{subsec:uncorrelatedPauli}
 
Consider now a Pauli channel in which the unitary operators acting on different subsystems are not correlated to each other. More specifically, we suppose that each qubit is acted on by a depolarizing channel with noise parameter $p$.
%i.e., $q_{\{m_in_i\}}$ are factorizable. For $(N+1)$-party d-dimensional system, the number of free parameters $q_{\{m_in_i\}}$ in the case of uncorrelated noise, given in Eq. (\ref{PCmulti}), are $(Nd^2 - 1)$, which is  12 for the three-qubit system and therefore, the evaluation of  the DC capacity, in this case is not easy to handle.
 %
%To study such noisy channels,  we restrict ourselves to the  depolarising channels with equal noise parameter $p$ which is an example of an uncorrelated channel. 
%If $\rho$ is a single qubit state, then the expression of depolarising channel is given in Eq. (\ref{Eq:De_ex}) with $\lambda_x = \lambda_y = \lambda_z = \frac{p}{3}$. 
  % Let us concentrate on the situation where each of the senders sends their parts individually to the receiver through a depolarising channel. Moreover we assume  that the p acting on all senders are same. 
Before analyzing the relation between the multiparty  DC capacity and quantum correlation measures, we compare the multiport dense coding capacities for the correlated channels with those of the uncorrelated ones. A three-qubit state, $\rho_{S_1S_2R}$, after the post-encoded qubits pass through independent (uncorrelated)  depolarizing channels, of equal strength, $p$, takes the form
\begin{eqnarray}
 &&\mathscr{D}(\rho_{S_1S_2R}) = (1-p)^2\rho_{S_1S_2R}\nonumber\\
  &+& \frac{(1-p)p}{3}\sum_{i=1}^3 (I_{S_1}\otimes\sigma^i_{S_2}\otimes I_R)\rho_{S_1 S_2 R}(I_{S_1}\otimes\sigma^i_{S_2}\otimes I_R) \nonumber \\
  &+& \frac{(1-p)p}{3}\sum_{i=1}^3 (\sigma^i_{S_1}\otimes I_{S_2}\otimes I_R)\rho_{S_1 S_2 R}(\sigma^i_{S_1}\otimes I_{S_2}\otimes I_R) \nonumber \\
  &+& \frac{p^2}{9}\sum_{i=j=1}^3 (\sigma^i_{S_1}\otimes\sigma^j_{S_2}\otimes I_R)\rho_{S_1 S_2 R}(\sigma^i_{S_1}\otimes\sigma^j_{S_2}\otimes I_R).\nonumber 
\end{eqnarray}
In the top panel of Fig. \ref{fig:corr_vs_Depolarising}, the capacity of DC is plotted against the total noise, $2H(p)$, of the uncorrelated channel, %parameter of the depolarizing channel, $p$, 
for various choices of $\alpha$ in the $g$GHZ state.  The bottom panel represents the DC capacity in the case of the fully correlated Pauli channel with respect to the noise, $H(\{q_i\})$, in this case, for the same $g$GHZ states.  
%the sum of the noise parameters $q_1 + q_2$, involved in the capacity of the $g$GHZ state, given in Eq. (\ref{Eq:gGHZ_noi_capa}).  
The amount of  correlated Pauli noise that can keep the $g$GHZ state dense codeable, is therefore higher than that of the  uncorrelated noise.
%, when the input state is the gGHZ state. 

\begin{figure}[t]
\begin{center}
\includegraphics[width=0.6\columnwidth,keepaspectratio,angle=270]{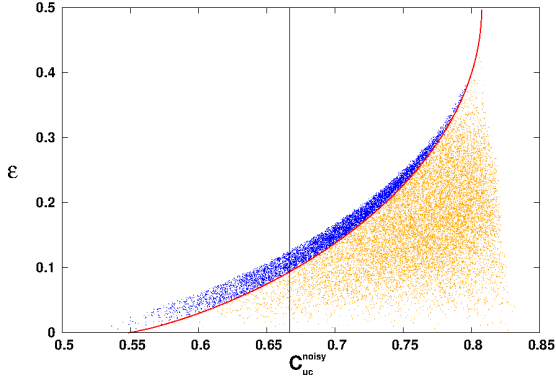} 
 \end{center}
\caption{(Color.) GGM vs. the raw DC capacity, $C_{uc}^{noisy}$, in the presence of the uncorrelated noise. Se text for further details. both axes represent dimensionless quantities. The vertical line at $C_{uc}^{noisy}=2/3$ again helps to read the actual capacity from the raw capacity.} \vspace{-1em}
% as depolarizing channel.  Like Pauli channel, there are set of states (blue dots) for which the eigenvalue in the receiver's side in GGM gives the maximum. Orange dots represent arbitrary three-qubit states. The $g$GHZ state (red line) and arbitrary three-qubit states are plotted with a very weak depolarizing noise parameter $p=0.04$.  All the quantities plotted are dimensionless. }
\label{fig:Depolarising_disco_GGM}
 \end{figure}

%We find from the Fig. \ref{fig:Depolarising_disco_GGM} that the Theorem 3 also holds for the depolarizing channel. Unlike Fig. \ref{fig:GGMcapacitynoise}, we plot GGM with respect to the capacity, \(C_{uc}^{noisy})\), for randomly generated three-qubit states.  Let us again consider the set of states for which the GGM and DC capacity are same for the noiseless case. In the presence of deplorizing channel, .......................  
To analyze the relation between the DC capacity and quantum correlation, we plot, in Fig. \ref{fig:Depolarising_disco_GGM}, the GGM against \(C^{noisy}_{uc}\), the DC capacity for two senders and a single receiver, with the post-encoded quantum systems being sent to the receiver via uncorrelated depolarizing channels, for arbitrary pure three-qubit states, which are numerically generated by choosing $5\times 10^4$ random states. 
%using the uniform Haar measure on the corresponding space. 
We choose the noise parameter, $p$,  as $0.04$ for the purpose of the figure (Fig. \ref{fig:Depolarising_disco_GGM}).  Fig. \ref{fig:corr_vs_Depolarising} shows that for small values of $p$, the $g$GHZ state remains dense codeable even for small values of \(\alpha\). In Fig. \ref{fig:Depolarising_disco_GGM}, The blue dots are the ones which satisfy condition $(ii)$ of Theorem 3. Note that condition $(i)$ is not well-defined in the current (uncorrelated) scenario. Most of them lie above the red curve of the $g$GHZ states. The remaining states are represented by orange dots.
%{\bf Arbitrary three-qubit states and the $g$GHZ state behave similarly like the Case 1 of fully correlated Pauli channel.  Blue dots in the  Fig. \ref{fig:Depolarising_disco_GGM} represent the set of states whose DC capacity and GGM were same as  $g$GHZ state in the case of noiseless channel. In particular, this is the set of states for which the maximum eigenvalue of GGM comes from the receiver's end. We also have numerically checked and found  that the arbitrary states to increase their dense coding capability need higher multipartite entanglement than that of the $g$GHZ state, in the presence of moderate depolarizing noise.}

%The set of states for which DC capacity and GGM are same with the gGHZ without noise, also show similar behavior with respect to DC capacity and GGM with the gGHZ state.
% However, the discord score shows only states which are below the curves of the gGHZ state in the plane of \(\delta_D\) and \(C^{noisy}_{uc}\). It is plausible that the states different from the gGHZ 
%{\bf Why discord score behave differently!!!!} 
% against GGM for a fixed amount of noise parameter, $p$ in Fig. 8. In this case, ....is there any relation with gGHZ state? 

% the noise parameter p for the gGHZ state. As expected becomes non dense codeable very fast with the increase of p. 
\vspace{-1em}
\section{Conclusion}
\label{sec:conclusion}

For transmission of classical information over noiseless and memory-less quantum channels, the capacity in the case of a single sender and a single receiver is well-studied. However, point-to-point communication is of limited commercial use and the exploration of quantum networks with multiple senders and receivers is therefore of far greater interest. Moreover, creation of multipartite systems with quantum coherence, 
%currently is one of the main aims in the laboratory and this is because  multiparticle systems are undoubtedly 
the essential ingredient for several quantum communication as well as computational tasks, is currently being actively pursued in laboratories around the globe. Establishment of connections between multipartite quantum correlation and capacities are usually hindered by the
%have, in principle, some obstacles   due to 
unavailability of a unique multiparty quantum correlation measure even for pure states, and the plethora of possibilities for multiparty communication protocols. 

For a communication scenario involving several senders and a single receiver, we establish the relation between capacities of classical information transmission and multipartite computable quantum correlation measures, both for the noiseless as well as noisy channels. We show that there are hierarchies among multipartite states according to the capacities of the dense coding protocol and hence obtain a tool to classify quantum states according to their usefulness in quantum dense coding.
% which depends on the amount of noise in that channel.  
The results can be an important step forward in building up communication networks using multipartite quantum correlated states in realizable systems.

 \vspace{-1em}
 \begin{acknowledgments}

RP acknowledges an INSPIRE-faculty position at the Harish-Chandra Research Institute (HRI) from the Department of Science and Technology, Government of India.
We acknowledge computations performed at the cluster computing facility at HRI. 
% (http://cluster.hri.res.in/).
\end{acknowledgments}

 \end{document}